\documentclass[a4paper,11pt,widemargins]{article}

\begin{document}

\title{A biology journal that can teach physicists a lesson in peer review\footnote{An edited version of this commentary 
has been published as: R.E. Goldstein, ``A biology journal provides a lesson in peer review'', {\it Physics Today} {\bf 69}(12), 10 (2016).}}

\author{Raymond E. Goldstein\footnote{R.E.Goldstein@damtp.cam.ac.uk}\\ 
Department of Applied Mathematics and Theoretical Physics, \\
Centre for Mathematical Sciences, University of Cambridge, \\ Wilberforce Road, Cambridge CB3 0WA, United Kingdom}

\date{\today}

\maketitle

After weeks of waiting, the email finally arrives from {\it Physical Review Z} with the reports on your paper 
``A theory of the nonuniform browning of toast''.  Hope turns to despair as you read the editor's cover letter saying 
that because the reviews contain a sufficiently strong criticism of your paper it cannot be accepted.  
Then you read the reports and discover that Referee A said ``This is a wonderful paper, full of interesting new results 
that will surely be of interest to a wide audience.  I am particularly impressed with Eq. 7 and its 
consequences and expect it to have broad applicability in physics.'' So what's the problem?  Well, 
Referee B said ``I fail to see any great significance in the results presented, and doubt the paper will be of broad interest.  
In addition, the result, Eq. 7, is wrong, calling into question the entirety of the subsequent results''.  
Besides wondering why the editor listened to Referee B and not A, or at least why he did not try to figure out whether 
Eq. 7 really does have a problem, this hypothetical scenario points to a deeper problem with the peer review system 
at most journals (including most physics journals):   With a few exceptions, there is no mechanism for the referees 
and the editor to discuss the paper and arrive at a consensus recommendation {\it before} reports are communicated to the authors.  
Instead, the initial recommendation is based on some sort of implicit averaging (in the editor's mind) of the reports 
and it is left up to the author to argue back from what might have been an error on the part of the referee, or more 
generally to somehow try to answer what are often conflicting and contradictory reports.  
Yes, it is true that in subsequent rounds of review the referees often see earlier reports, but again there is 
no consensus-building process.  All of us who spend our professional lives dealing (on both sides) with this deeply flawed process deserve better.

Remarkably, this problem has been solved by a relatively new, high-profile, open access journal in the life sciences: 
{\it eLife}.  Launched in 2012, the journal is a joint effort between the Max Planck Society in Germany, the Wellcome 
Trust in the UK, and the Howard Hughes Medical Institute in the US, and is meant as a direct challenge to the 
troika of journals ({\it Science}, {\it Nature}, and {\it Cell}) that dominates the life sciences and, in the case of the first 
two, also has a strong presence in certain areas of physics.  The editorial decision-making process in {\it eLife}
is vastly different from those three journals, which mostly rely on in-house editors, rather than practicing 
scientists, to do an initial cull of submissions before sending only a minority out to review.  The main point of 
interest here is the reviewing process at {\it eLife}, which I highlight in the hope that physics journals might adopt these practices.  
I come to this point of view having first learned about the journal from colleagues connected with the Wellcome Trust, 
from which I have long-term funding, and having now published several articles in {\it eLife} \cite{1,2,3}  and also acted as a referee for it, 
but I have no formal association with the journal.  I have spoken with many, many people about the review process in {\it eLife}
and can report that even those whose papers were ultimately rejected spoke highly of it!

The essence of the {\it eLife} review process is an online discussion between the referees and the handling editor of a paper to arrive 
at a single consensus report (a “decision letter”) that is sent to the authors.  The discussion takes the form of a series of 
entries in a chain of postings behind the journal's firewall, where the identities of the referees and editor are all known to each other. 
The original reports of the individual referees are not sent to the authors, although comments they contain may of course be 
incorporated into the consensus report where appropriate.  The decision letter lays out the status of the paper - whether 
minor changes are needed, a major revision is warranted, or the paper is simply rejected - and in the first two cases, describes 
what needs to be done to make the paper acceptable for publication.  This letter, and the authors' reply to it, are both published 
along with the paper if it is ultimately accepted.  If a paper is rejected, then the process is like that of most journals in that 
the editor typically sends a summary statement along with the full reports so the authors see all the concerns raised.

The advantages of this system are obvious.  If the referees differ on a technical point then at the very least they discuss 
it and arrive at a single point of view (e.g., clarification is needed, or there is a problem that needs addressing).   {\it eLife}
has very high standards with regard to importance, broad interest, and novelty, and here too if there is an issue the referees 
and editor speak with one voice to the authors.  The fact that the reviewers are known to each other online naturally tends to 
enforce both higher standards and greater civility than would be the case with anonymity.  Because there is a single decision 
letter, the authors spend less time doing additional experiments or responding to contradictory reports. 

What are the disadvantages?  It can certainly take more time to conduct the online discussion.  It requires more up-front work 
from the editors, who must coordinate the process.  I would suggest that ultimately less work is involved in subsequent reviewing 
rounds as the editor can usually decide on the suitability of the revised paper without sending it back to the referees.  
The process certainly requires more work from referees, who must engage in the online discussion, but will likely do a better job 
precisely because of that.  But ultimately, the process is much more satisfying to all involved.

So, I leave it as a challenge to physics journals to adopt this review process.  I have enough experience with the 
editorial issues confronted by physics journals to know that such a change would involve a considerable amount of work.  
So let us start with a single journal, say, {\it Physical Review Letters}, and see if we can make the review process work better.

For discussions on this subject and feedback on drafts this article I am grateful to Randy Schekman and Andy Collins, 
respectively Editor-in-Chief and Executive Editor of {\it eLife}, and to my colleague Eric Lauga.   This work was supported 
in part by a Senior Investigator Award from the Wellcome Trust and an Established Career Fellowship from the Engineering and 
Physical Sciences Research Council.

\end{document}